# Density of states near an Anderson transition in a four-dimensional space. Renormalizable models

I. M. Suslov

*P. L. Kapitsa Institute of Physical Problems, Russian Academy of Sciences, 117334 Moscow, Russia*
(Submitted 6 June 1996)
Zh. Éksp. Teor. Fiz. **111**, 220–249 (January 1997)

Asymptotically accurate results are obtained for the average Green function and density of states of a disordered system for a renormalizable class of models (as opposed to the lattice model examined previously [I. M. Suslov, Zh. Éksp. Teor. Fiz. **106**, 560 (1994)]. For $N \sim 1$ (where $N$ is the order of perturbation theory), only the parquet terms corresponding to the higher powers of large logarithms are taken into account. For large $N$, this approximation is inadequate because of the higher rate of increase with respect to $N$ of the coefficients for the lower powers of the logarithms. The latter coefficients are determined from the renormalization condition for the theory expressed in the form of a Callan–Symanzik equation using the Lipatov asymptote as boundary conditions. For calculating the self-energy at finite momentum, a modification of the parquet approximation, is used that allows the calculations to be done in an arbitrary finite logarithmic approximation, including the principal asymptote in $N$ of the expansion coefficients. It is shown that the phase transition point moves in the complex plane, thereby ensuring regularity of the density of states for all energies and avoiding the ''false'' pole in such a way that the effective interaction remains logarithmically weak. © *1997 American Institute of Physics.* [S1063-7761(97)01401-7]

## 1. INTRODUCTION

The problem of calculating the average Green function that determines the density of states for the Schrödinger equation with a gaussian random potential is mathematically equivalent to the problem of a second-order phase transition with an $n$-component order parameter $\varphi = (\varphi_1, \varphi_2, \ldots, \varphi_n)$ in the limit $n \to 0$;[1,2] the coefficients in the Ginzburg–Landau Hamiltonian

$$H\{\varphi\} = \int d^d x \left\{ \frac{1}{2} c |\nabla \varphi|^2 + \frac{1}{2} \kappa_0^2 |\varphi|^2 + \frac{1}{4} g_0 |\varphi|^4 \right\} \quad (1)$$

are related to the parameters of a disordered system by the equations

$$c = 1/2m, \quad \kappa_0^2 = -E, \quad g_0 = -W^2 a_0^d / 2, \quad (2)$$

where $d$ is the dimensionality of the space, $m$ is the particle mass, $E$ is the energy relative to the lower boundary of the seed spectrum, $W$ is the amplitude of the random potential, and $a_0$ is the lattice constant. (In the following $c = 1$ and $a_0 = 1$.) The ''wrong'' sign on the coefficient of $|\varphi|^4$ leads to the ''false'' pole problem[3] and for a long time it was doubted that an $\varepsilon$-expansion could be constructed near a spatial dimensionality of $d = 4$.[4] Encouraging results in this area have been obtained recently by the author.[5,6]

It has been shown[6] that there are two fundamentally different classes of models which show up in estimates based on the optimal fluctuation method.[7,8] The probability $P(E,R)$ of the appearance of an energy level $E < 0$ owing to a fluctuation in a potential with characteristic size $R$ has the form

$$P(E,R) \sim \exp\{-S(E,R)\}, \quad (3)$$

where $S(E,R) \sim W^{-2} \gg 1$. The total probability $P(E)$ of the resulting level $E$, which determines the density of states $\nu(E)$, is obtained by integrating Eq. (3) with respect to $R$, which in the approximation of the saddle-point method reduces to replacing $R$ by $R_0$, the minimum point for $S(E,R)$. For $d < 4$ and $d > 4$ we have $R_0 \sim |E|^{-1/2}$ and $R_0 \sim a_0$, respectively.[5,6] For $d = 4$ (Fig. 1), the function $S(E,R) = \text{const} = S_0$, and the situation is close to degeneracy: for large $R$ the degeneracy is removed owing to the finiteness of $E$ and $S(E,R) - S_0 \sim E^2 R^d$, while for small $R$ the deviation of the spectrum $\varepsilon(k)$ from quadratic is large. If $\varepsilon(k) = k^2 + \beta k^4$, then for $\beta > 0$ the function $S(E,R)$ lies above $S_0$, ensuring the appearance of a minimum at $R_0 \sim |E|^{-1/4}$, while for $\beta < 0$ it lies below and the minimum is attained at $R_0 \sim a_0$;[6] thus, models with $\beta > 0$ and $\beta < 0$ yield a different asymptote for the fluctuation tail as $E \to -\infty$. For small negative $E$ the boundary between the two types of models shifts and is no longer sharp, so that integrating Eq. (3) with respect to $R$ results in a competition between the contributions from the minimum $S_1$ and the higher lying plateau $S(E,R) = S_0$, whose width increases without bound as $|E|$ is reduced:

$$P(E) \sim \nu(E) \sim e^{-S_1} + \left(\frac{J}{|E|}\right)^\alpha e^{-S_0}, \quad (4)$$

where $J \sim 1/m a_0^2 \sim 1$. As $S_1$ is increased the second term (the contribution of the plateau) becomes dominant before $S_1$ approaches $S_0$. Direct integration of Eq. (3) with respect to $R$ yields an exponent $\alpha = 1/2$,[6] which cannot be taken seriously, since the accuracy of the method does not allow for an estimate of the coefficient of the second exponential. The exact value of $\alpha$ is 1/3 (see below).

In the domain of applicability of the optimal fluctuation method, the damping $\Gamma$, defined by the imaginary part of the self-energy $\Sigma(p,\kappa)$ for $p=0$ ($\kappa$ is the renormalized value of $\kappa_0$), is proportional to the density of states $\nu(E)$ and, when the dimensionality is taken into account, can be estimated as

$$\Gamma \sim J\left\{e^{-S_1} + \left(\frac{J}{|E|}\right)^{1/3} e^{-S_0}\right\}. \quad (5)$$

The energy always enters in the combination $E+i\Gamma$, and in the neighborhood of an Anderson transition $|E|$ can be replaced by $\Gamma$. It is easily verified that the first term in brackets is dominant when $S_1 < 3S_0/4$, and the second when the inequality is reversed. Since $S(E,R) \sim W^{-2}$,[6] in the limit of weak disorder a sharp boundary $S_c = 3s_0/4$ appears between the two types of models: for $S_1 < S_c$ the optimum fluctuation is determined by the atomic scale length and the discreteness of the lattice is of fundamental importance, by analogy with the case $d > 4$. For $S_1 > S_c$ fluctuations with a large radius are important and the analysis can be carried out in a continuum model with a quadratic spectrum; the situation is analogous to that for the lowest dimensionalities.

The above classification of models is directly related to the renormalizability of the theory. The $N$th-order graph for the self-energy $\Sigma$ has momentum dimensionality $k^r$, where $r=2+(d-4)N$. For $d>4$ the degree of divergence at high momenta increases with the graph order, and the theory is unrenormalizable;[9] a cutoff parameter $\Lambda$ must be introduced explicitly as an indication of the significance of the structure of the Hamiltonian on an atomic scale. For $d<4$ we have $r<2$ at all $N$: when its value is deducted from each graph for $p=\kappa=0$ the index $r$ is reduced by 2 and the difference $\Sigma(p,\kappa) - \Sigma(0,0)$ contains no divergences, which are absorbed by $\Sigma(0,0)$, and leads only to a shift in the energy origin. For $d=4$ the difference $\Sigma(p,\kappa) - \Sigma(0,0)$ contains logarithmic divergences which are removed by renormalizing the charge and Green function;[9,10] however, it is necessary to keep in mind that in the standard proofs of renormalizability only distances greater than $\Lambda^{-1}$ are considered. Of course, scale lengths shorter than $\Lambda^{-1}$ do not make the $\delta$-function contributions that are so important for $\Lambda \to \infty$. The above estimate shows that this is not always so: the renormalizable contribution from large distances (the contribution of the plateau) is dominant only for $S_1 > S_c$; otherwise, it is small compared to the unrenormalizable contribution from small distances.

Therefore, there are four fundamentally different types of theory: (1) an unrenormalizable theory for $d>4$; (b) unrenormalizable theories under logarithmic conditions ($d=4$, $S_1 < S_c$); (c) renormalizable theories under logarithmic conditions ($d=4$, $S_1 > S_c$); and (d) theories that are renormalizable with a single subtraction (superrenormalizable) for $d<4$. Cases (a) and (b) have been examined in Refs. 5 and 6, respectively. In this paper we examine case (c), the zeroth approximation for the $4-\varepsilon$ theory, which belongs to type (d).

The exponent $\alpha$ in Eq. (4) can be determined from the renormalization condition for the plateau contribution. The contribution of the latter to the damping $\Gamma$, which depends on $\Lambda$ and the bare values for $\kappa_0$ and $g_0$, becomes a function of $\kappa$ and $g$ alone upon transforming to the renormalized quantities. From dimensional considerations $\Gamma = \kappa^2 f(g)$, where the function $f(g)$ is determined mainly by the exponential $\exp(-1/ag)$ owing to the need to agree with the result from the optimal fluctuation method as $E \to -\infty$, in which $g \approx g_0$. Given the relationship between the renormalized and bare charges,[11]

$$g = \frac{g_0}{1 + W_2 g_0 \ln(\Lambda/\kappa)}, \quad W_2 = K_4(n+8) \quad (6)$$

($K_4 = (8\pi^2)^{-1}$ is the area of a unit sphere in four-dimensional space, divided by $(2\pi)^4$), we have

$$\Gamma \sim \kappa^2 \exp\left\{-\frac{1}{ag_0} - \frac{W_2}{a} \ln \frac{\Lambda}{\kappa}\right\}$$
$$\sim \Lambda^2 \left(\frac{\Lambda^2}{\kappa^2}\right)^{-W_2/2a - 1} \exp\left(-\frac{1}{ag_0}\right), \quad (7)$$

which, given that $J \sim \Lambda^2$, $\kappa^2 = |E|$,[1)] and $a = -3/8\pi^2$, reproduces the second term of Eq. (5). This value of $a$ is obtained by a method[3,12,13] employing the standard instanton solution for $d=4$.[14]

The need to correctly account for the factorial divergence of a number of perturbation theories examined via Lipatov's method,[14] according to which subsequent coefficients in the expansion in $g_0$ are determined by saddle-point configurations, i.e., instantons, of the corresponding functional integrals, has been clarified previously.[5,6] An instanton in the Lipatov method satisfies the same equation as a typical wave function in the field of an optimal fluctuation (see Chapter IV of Ref. 8). In this way, the model classification given above shows up in yet another fundamental guise–the divergence of a number of perturbation theories. Unlike the lattice models for $d \geq 4$,[6] the continuum models for $d<4$,[15] and the renormalizable massless theories,[14,16] applying the Lipatov method to studies of four-dimensional models (1) with $\kappa \neq 0$ requires that certain difficulties associated with the absence of ''true'' instantons (Sec. 7) be overcome.

In order to obtain asymptotically accurate (in the limit of weak disorder) results in four-dimensional lattice models,[6] in the expansion

$$\Sigma(0,\kappa) - \Sigma(0,0) = \kappa^2 \sum_{N=1}^{\infty} g_0^N \sum_{K=0}^{N} A_N^K \left(\ln \frac{\Lambda}{\kappa}\right)^K \quad (8)$$

it is necessary to include: (a) the parquet coefficients $A_N^N$ corresponding to the principal logarithmic approximation, and (b) for $N \geq N_0 \gg 1$, the coefficients $A_N^0$ and $A_N^1$, which have the maximum growth rate with respect to $N$ and dominate the higher orders of perturbation theory. They yield a nonperturbative contribution, which is related to the divergence of the series and does not depend on the choice of $N_0$. The qualitative result consists of a shift in the transition point from the real axis into the complex plane, which leads to regularity of $\nu(E)$ in the neighborhood of an Anderson transition and elimination of the false pole. This approximation ''deteriorates'' as $S_1$ approaches $S_c$:[6] (a) the equation for $\Gamma(E)$ has physically meaningless solutions when $S_1 > S_c$; (b) the contribution from the approximations fol-lowing the principal logarithmic approximation, which are determined by coefficients $A_N^{N-K}$ with $K \sim 1$, increases rapidly as $S_1 \to S_c$; and, (c) the plateau (Fig. 1) makes a contribution whose strong energy dependence indicates a growing role for the coefficients $A_N^K$ with $K \neq 0$ (in the lattice models the weak dependence of $S_1$ on $E$ makes the zero-logarithmic contribution predominate) that becomes important for $S_1 \approx S_c$. Thus, if the "highest" and "lowest"-order logarithms are dominant in the lattice models, then, in general, on going to the renormalizable models the contributions from all $K$ become important in the sum (8).

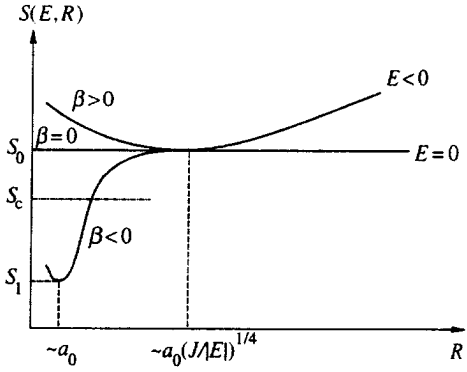

FIG. 1. The dependence of $S(E,R)$ on $R$ for $E=\text{const}$ when $d=4$.

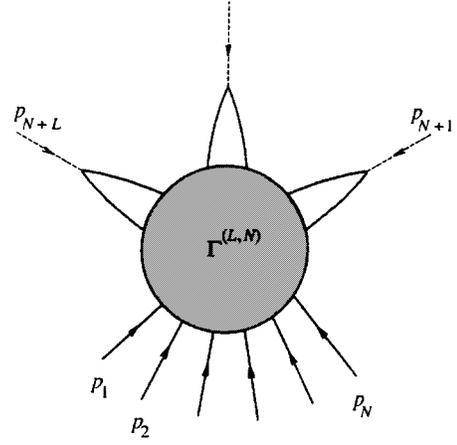

FIG. 2. The vertex $\Gamma^{(L,N)}$ with $N$ free ends and $L$ closed two-line loops studied in the renormalization theory of Ref. 10.

In the latter case we arrive at the following statement of the problem: let us choose an integer $N_0$ that is large compared to 1, but small compared to the large parameters of the theory. For $N < N_0$ we retain only the parquet coefficients $A_N^N$, which are distinguished by large logarithms, in Eq. (8). For $N \gtrsim N_0$, in general, all the terms are important in the sum over $K$, but the condition $N \gg 1$ allows us to calculate the coefficients $A_N^K$ in the principal asymptotic dependence on $N$. The latter problem is solved in the following way: the renormalizability of the theory, expressed in the form of the Callan–Symanzik equation (Sec. 2), leads to a system of equations for the $A_N^K$ that determines the coefficients with $K \neq 0$ in terms of specified $A_N^0$. On the other hand, the Lipatov method reproduces the coefficients $A_N^K$ with small $K$ well, so that they can be used as boundary conditions for this system of equations. In this way it is possible to determine all $A_N^K$ with $N \gg 1$ (Sec. 3), which for small $g_0$ enables us to find the sum in Eq. (8) and determine the energy dependence of the damping $\Gamma$ (Sec. 4).

For calculating the density of states (Sec. 6) it is necessary to find the self-energy $\Sigma(p,\kappa)$ for finite momenta, which requires solving of the parquet equations in the principal logarithmic approximation.[6] In the present theory $\Sigma(p,\kappa)$ is the sum of a nonperturbative contribution, mainly determined by the Lipatov asymptote, and a quasi-parquet contribution corresponding to a logarithmic approximation of arbitrary finite order, which allows only for the principal asymptote behavior in $N$. The calculations in the next order can be carried out with the aid of a curious modification of the parquet approximation (Sec. 5).

## 2. SYSTEM OF EQUATIONS FOR THE COEFFICIENTS $A_N^K$

In the following we shall only be interested in logarithmic divergences, assuming the quadratic to be eliminated by the renormalization $\kappa$. The Callan–Symanzik equations can be derived in the usual way,[10] but we need them in a somewhat nontraditional form. Dimensional considerations imply that a vertex $\Gamma^{(L,N)}$ with $N$ free ends and $L$ two-line loops (Fig. 2) can be written in the form[10]

$$\Gamma^{(L,N)}(p_i;\kappa,g_0,\Lambda) = \kappa^{d-N(d-2)/2-2L}$$
$$\times \widetilde{\Gamma}^{(L,N)}(p_i/\kappa;g_0,\Lambda/\kappa), \quad (9)$$

which allows us to proceed to examining $\widetilde{\Gamma}^{L,N}$ (we omit the tilde in the following). Assuming that the bare charge $g_0$ is a function of $\Lambda$ and introducing the renormalized charge $g_\mu$ applicable to a scale $\mu \gg \kappa$, in view of the multiplicative renormalizability of $\Gamma^{L,N}$,[10] we have

$$\Gamma_R^{(L,N)}\left(\frac{p_i}{\kappa};g_\mu,\frac{\mu}{\kappa}\right) = Z^{N/2}\left(\frac{Z_2}{Z}\right)^L \Gamma^{(L,N)}\left(\frac{p_i}{\kappa};g_0,\frac{\Lambda}{\kappa}\right), \quad (10)$$

where $Z$ and $Z_2$ are functions of $g_0$ and $\Lambda/\mu$. Since $\Gamma_R^{(L,N)}$ is independent of $\Lambda$, we have

$$d\Gamma_R^{(L,N)}/d\ln\Lambda = 0,$$

which after substitution of Eq. (10) gives the Callan–Symanzik equation:

$$\left[\frac{\partial}{\partial \ln \Lambda} + W(g_0)\frac{\partial}{\partial g_0} + \left(L - \frac{N}{2}\right)\eta(g_0)\right.$$
$$\left. - L\eta_2(g_0)\right]\Gamma^{(L,N)}\left(\frac{p_i}{\kappa};g_0,\frac{\Lambda}{\kappa}\right) = 0. \quad (11)$$

The Gell-Mann–Low function $W(g_0)$ and the scaling functions[10] $\eta(g_0)$ and $\eta_2(g_0)$ are defined by the equations

$$W(g_0) = \frac{dg_0}{d\ln\Lambda}, \quad \eta(g_0) = -\frac{d\ln Z}{d\ln\Lambda},$$

$$\eta_2(g_0) = -\frac{d\ln Z_2}{d\ln\Lambda} \quad (12)$$

and depend *a priori* on $\Lambda/\mu$; however, by writing the three Eqs. (11) for different $L$ and $N$, expressing $W$, $\eta$, and $\eta_2$ in terms of $\Gamma^{(L,N)}$, and noting that the latter functions are independent of the arbitrary parameter $\mu$, it is easy to confirm that the first of them are independent of $\Lambda/\mu$.

In order to find the renormalization law for the self-energy we use the Ward identity for the Green function $G(p,\kappa)$:

$$\frac{\partial G^{-1}(0,\kappa)}{\partial \kappa_0^2} = \frac{\partial \kappa^2}{\partial \kappa_0^2} = \Gamma^{(1,2)}(\kappa), \quad (13)$$

which, when integrated, yields

$$\kappa_0^2 - \kappa_c^2 = \kappa^2 Y\left(g_0, \frac{\Lambda}{\kappa}\right) \equiv \int_0^{\kappa^2} d\kappa^2 [\Gamma^{(1,2)}(\kappa)]^{-1}, \quad (14)$$

where $\Gamma^{(1,2)}(\kappa) \equiv \Gamma^{(1,2)}(p_i=0, g_0, \Lambda/\kappa)$ and $\kappa_c^2 = \Sigma(0,0)$. The function $Y$ satisfies the equation

$$\left[\frac{\partial}{\partial \ln \Lambda} + W(g_0) \frac{\partial}{\partial g_0} + V(g_0)\right] Y\left(g_0, \frac{\Lambda}{\kappa}\right) = 0 \quad (15)$$

with $V(g_0) \equiv \eta_2(g_0)$, which is easily confirmed by applying the operator in square brackets to Eq. (14) and using Eq. (11). Given that $\kappa_0^2 = \kappa^2 + \Sigma(0,\kappa)$, Eq. (14) can be rewritten in the form

$$\kappa^2 + \Sigma(\kappa,0) - \Sigma(0,0) = \kappa^2 Y(g_0, \Lambda/\kappa) \quad (16)$$

and a comparison with Eq. (8) yields the following logarithmic expansion for $Y$:

$$Y\left(g_0, \frac{\Lambda}{\kappa}\right) = \sum_{N=0}^{\infty} g_0^N \sum_{K=0}^{N} A_N^K \left(\ln \frac{\Lambda}{\kappa}\right)^K, \quad (17)$$

with $A_0^0 = 1$. Expanding the functions $W$ and $V$ in the series

$$W(g_0) = \sum_{N=2}^{\infty} W_N g_0^N, \quad V(g_0) = \sum_{N=1}^{\infty} V_N g_0^N, \quad (18)$$

whose leading coefficients[10] are[2)]

$$W_2 = K_4(n+8), \quad W_3 = -K_4^2(9n+42),$$
$$V_1 = -K_4(n+2), \quad V_2 = 3K_4^2(n+2), \quad (19)$$

substituting Eqs. (17) and (18) into Eq. (15), and collecting terms with the same powers of $g_0$ and the logarithms, we obtain a system of equations for the coefficients $A_N^K$:

$$-KA_N^K = \sum_{M=1}^{N-K+1} [W_{M+1}(N-M) + V_M] A_{N-M}^{K-1},$$
$$K = 1, 2, \ldots, N. \quad (20)$$

## 3. A STUDY OF THE COEFFICIENTS $A_N^K$

Equation (20) is a recurrence relation that determines the $A_N^K$ in terms of specified $A_{N-1}^{K-1}, A_{N-2}^{K-1}, \ldots, A_{K-1}^{K-1}$ and can be used to express all the $A_N^K$ in terms of a single sequence $A_N^0$. The coefficients $W_N$ and $V_N$ can be determined from Eq. (20) if we specify two sequences $A_N^1$ and $A_N^2$ in addition to $A_N^0$. Thus, the renormalizability of the theory sharply reduces the arbitrariness in the choice of coefficients in Eq. (8).

Information on the coefficients $A_N^K$ with $N \gg 1$ can be obtained by the Lipatov method. According to Sec. 7, the $N$th order contribution to $\Sigma(0,\kappa)$ has the form

$$\kappa^2 g_0^N c_2 \Gamma(N+b) a^N (\ln N)^{-\gamma} \exp\left(\sigma \ln \frac{\Lambda}{\kappa}\right), \quad (21)$$

where

$$b = \frac{n+8}{3}, \quad a = -3K_4, \quad \gamma = \frac{n+2}{6}, \quad \sigma = \frac{n+8}{3}. \quad (22)$$

Comparing this with the expansion (8), we obtain

$$A_N^K = \frac{\sigma^K}{K!} A_N^0, \quad A_N^0 = c_2 \Gamma(N+b) a^N (\ln N)^{-\gamma}. \quad (23)$$

Whereas in the lattice models[6] the Lipatov asymptote only reproduces the zero-logarithmic and first-logarithmic contributions, here it yields some ''extra'' logarithms. Formally, in Eq. (23) $K = 0, 1, \ldots, \infty$, while in Eq. (8) $K \leq N$. The reason for this is the rapid drop in $A_N^K$ with increasing $K$ and the limited accuracy ($\sim 1/N$) of the leading asymptote. The result (23) can be believed only for small $K$, but this is enough to use it as a boundary condition for the system of Eqs. (20).

Writing out Eq. (20) for small $K$,

$$-1 \cdot A_N^1 = [W_2(N-1) + V_1] A_{N-1}^0 + [W_3(N-2)$$
$$+ V_2] A_{N-2}^0 + \ldots + [W_N \cdot 1 + V_{N-1}] A_1^0 + V_N A_0^0,$$

$$-2 \cdot A_N^2 = [W_2(N-1) + V_1] A_{N-1}^1 + [W_3(N-2)$$
$$+ V_2] A_{N-2}^1 + \ldots + [W_N \cdot 1 + V_{N-1}] A_1^1 \quad (24)$$

. . . . . . . .

and assuming that Eq. (23) is valid for $A_N^0$, it is easy to confirm the factorial growth in $N$ for all $A_N^K$ with $K \sim 1$. Retaining only the first terms in leading order in $N$ on the right-hand sides of Eqs. (24), we obtain

$$A_N^K = \frac{N!}{K!(N-K)!} (-W_2)^K A_{N-K}^0 \xrightarrow{K \ll N} \frac{1}{K!}$$
$$\times \left(-\frac{W_2}{a}\right)^K A_N^0, \quad (25)$$

which, given that $\sigma = -W_2/a$ (see Eqs. (19) and (22)), reproduces the result (23) for $K \neq 0$ and establishes its domain of applicability, $K \ll N$. Retaining only the first terms on the right of Eqs. (24) is justified when $W_N$ and $V_N$ increase more slowly than $A_N^0$,[3)] which may regarded as a consequence of the validity of Eq. (23) for $K = 0, 1, 2$.

For $K$ close to $N$ and assuming that $x_N = A_N^N$ and $y_N = A_N^{N-1}, \ldots$, we obtain a system of difference equations from Eq. (20),

$$-Nx_N = [W_2(N-1) + V_1] x_{N-1},$$
$$-(N-1) y_N = [W_2(N-1) + V_1] y_{N-1} + [W_3(N-2)$$
$$+ V_2] x_{N-2}, \quad (26)$$

. . . . . . . .

which can be solved by the method of variation of constants and used to successively determine $A_N^N, A_N^{N-1}, \ldots$. For the parquet coefficients we have

$$A_N^N = (-W_2)^N \frac{\Gamma(N-\beta)}{\Gamma(N+1)\Gamma(-\beta)}, \quad \beta = -\frac{V_1}{W_2} = \frac{n+2}{n+8} \tag{27}$$

in agreement with Ginzburg[11] (see Ref. 6). For $A_N^{N-K}$ with $K \sim 1$ it is easy to identify the leading asymptote in $N$ and prove the following result by induction:

$$A_N^{N-K} = \frac{1}{K!} \left( -\frac{W_3}{W_2^2} N \ln N \right)^K A_N^N. \tag{28}$$

In order to study the $A_N^K$ with arbitrary $K$ we use the estimate $A_{N-1}^K / A_N^K \lesssim 1/N$, which is valid for Eqs. (23) and (28) and is confirmed by the result for all $K$. Retaining the two leading terms in $N$ on the right of Eq. (20), we have

$$-KA_N^K = [W_2(N-1)+V_1]A_{N-1}^{K-1} + W_3 N A_{N-2}^{K-1},$$
$$K = 1,2,\ldots,N. \tag{29}$$

The principal term in $N$ is not sufficient, since the calculation of arbitrary $A_N^K$ from known $A_N^0$ requires $\sim N$ iterations, which for an accuracy $\sim 1/N$ in each iteration leads to a buildup of errors. The last terms in Eq. (24), which contain $W_N$ and $V_N$, generally give corrections $\sim 1/N$, but are present only in the equations with $K=1, 2$ and do not lead to an accumulation of errors. We assume by definition that $A_N^{N+1} = 0$, which accounts for the absence of the latter term in the equation with $K=N$. Making the substitution

$$A_N^K = (-W_2)^K \frac{\Gamma(N-\beta)}{\Gamma(K+1)\Gamma(N-K-\beta)} A_{N-K}^0 X_{N,N-K} \tag{30}$$

and noting that $X_{N+1,M} - X_{N,M} \sim 1/N$, we arrive at the equation

$$X_{N,M} = X_{N-1,M} + \frac{f(M)}{N} X_{N-1,M-1} \tag{31}$$

with the boundary conditions

$$X_{NN} = 1, \quad X_{N0} = 1, \tag{32}$$

where the function $f(M)$ is defined by

$$f(M) = \frac{W_3}{W_2} (M-1-\beta) \frac{A_{M-1}^0}{A_M^0}. \tag{33}$$

Equation (31) is convenient for studying a problem with initial conditions

$$X_{0,M} = \phi_M, \quad \text{where} \quad \phi_M = 0 \quad \text{for} \quad M = -1, -2, \ldots, \tag{34}$$

where $\phi_M$ can be chosen so as to satisfy the boundary conditions (32). Iteration of Eq. (31) yields

$$X_{NM} = \phi_M + B_N^1 f(M) \phi_{M-1} + B_N^2 f(M) f(M-1) \phi_{M-2}$$
$$+ \ldots + B_N^N f(M) f(M-1) \ldots f(M-N+1) \phi_{M-N}, \tag{35}$$

where the coefficients $B_N^K$ are given by the sum of the $C_N^K$ terms,

$$B_N^K = \sum_{\{p_i\}} \frac{1}{(1+p_1)(1+p_2)\ldots(1+p_K)}, \tag{36}$$

while $p_1, p_2, \ldots, p_K$ is a selection without replacement from the sequence $0, 1, \ldots, N-1$. For $K \ll N$, sampling with and without replacement are essentiallly equivalent, and we obtain

$$B_N^K \approx \frac{(\ln N)^K}{K!}, \quad K \ll N. \tag{37}$$

In fact, for large $N$, the sum in (35) is always dominated by $K \ll N$, and noting that $\phi_M = 0$ for $M < 0$, we obtain

$$X_{NM} = \sum_{K=0}^{M} \frac{(\ln N)^K}{K!}$$
$$\times f(M) f(M-1) \ldots f(M-K+1) \phi_{M-K}. \tag{38}$$

For the product in (38) we have

$$f(M) f(M-1) \ldots f(M-K+1)$$
$$= \left( \frac{W_3}{W_2} \right)^K \frac{A_{M-K}^0}{A_M^0} \frac{\Gamma(M-\beta)}{\Gamma(M-K-\beta)} \xrightarrow{M \gg K} (f_\infty)^K, \tag{39}$$

where

$$f_\infty = \lim_{M \to \infty} f(M) = \frac{W_3}{aW_2} = \frac{3n+14}{n+8}. \tag{40}$$

For $M=0$ the boundary condition (32) gives $\phi_0 = 1$ and for $M = N \gg 1$ the sum in Eq. (38) is replaced by an integral which is calculated by the saddle-point method and, on comparison with the boundary conditions (32), determines $\phi_M$ for $M \gg 1$:

$$\phi_M = \begin{cases} 1, & M=0 \\ M^{-f_\infty}, & M \gg 1 \end{cases}. \tag{41}$$

Substituting Eq. (41) in Eq. (38) leads to the results ($M_0 = f_\infty \ln N$)

$$X_{NM} = \frac{(\ln N)^M}{M!} \left( \frac{W_3}{W_2} \right)^M \frac{\Gamma(M-\beta)}{\Gamma(-\beta) A_M^0} \quad \text{for} \quad M \ll \ln N, \tag{42}$$

$$X_{NM} = \frac{(\ln M)^\gamma}{M^{b+\beta}} \frac{e^{M_0}}{\sqrt{2\pi M_0}} \int_0^\infty dx \exp\left( -\frac{(M-M_0-x)^2}{2M_0} \right)$$
$$\times (\ln x)^{-\gamma} x^{b+\beta-f_\infty}$$
$$\text{for} \quad M_0 - M \ll M_0 \quad \text{or} \quad M > M_0 \tag{43}$$

for $M_0 - M \ll M_0$ or $M > M_0$. (In the first case the sum in Eq. (38) is determined by the term with $K=M$ and in the second, it is replaced by an integral.) Substituting Eq. (42) in Eq. (30) reproduces the result (28) and establishes its domain of applicability, $K \ll \ln N$. In the region $M \sim (M_0 - M)$, which is not described by Eqs. (42) and (43), the magnitude of $X_{N,M}$ is determined by the values of $\phi_M$ for $M \sim 1$, which, in turn, are determined by the coefficients $A_N^0$ with $N \sim 1$.

The latter can be determined to within a few percent by matching the first orders of perturbation theory with the Lipatov asymptote (see the examples in Refs. 14 and 15).

## 4. SUMMING THE SERIES FOR $\Sigma(0,\kappa)$

The above information on the coefficients $A_N^K$ can be used to identify the regions in the $(N,K)$ plane which make significant contributions to the sum (8). Region *1* of Fig. 3 is a nonuniversality region in which the behavior of the $A_N^K$ depends substantially on the specific values of $A_N^0$ with $N \sim 1$. Above and below region 1, the universal asymptotes (42) and (43), determined by the trivial coefficient $A_0^0 = 1$ and the Lipatov asymptote for $A_N^0$, respectively, are valid. The dashed line *2* denotes the saddle-point values of $K$ for $N = \text{const}$ assuming $|g_0| \ll 1$ and $g_0 \ln(\Lambda/\kappa) \sim 1$. When $N$ is reduced the saddle point vanishes and the parquet coefficients $A_N^N$ on the principal diagonal become dominant. An important contribution to the sum (8) comes from regions *3* and *4* which are adjacent to the dotted saddle-point line.

Region *3* gives a quasiparquet contribution, determined by coefficients $A_N^{N-K}$ with $K \sim 1$, for which Eq. (28) is valid:

$$\left[ Y\!\left(g_0, \frac{\Lambda}{\kappa}\right) \right]_{\text{quasiparq}} = \left[ \Delta + \frac{W_3}{W_2} g_0 \ln \Delta \right]^\beta,$$

$$\Delta = 1 + W_2 g_0 \ln \frac{\Lambda}{\kappa}. \tag{44}$$

To within logarithmic accuracy the quantity $\Delta$ in the logarithm can be replaced by its minimum value $\widetilde{\Delta} \sim |g_0| \ln |g_0|$ (see below), since for $\Delta \gg \widetilde{\Delta}$ the logarithmic term is unimportant and Eq. (44) can be rewritten in the form

$$\left[ Y\!\left(g_0, \frac{\Lambda}{\kappa}\right) \right]_{\text{quasiparq}} \approx \left[ 1 + W_2 g_1 \ln \frac{\Lambda}{\kappa} \right]^\beta, \tag{45}$$

$$g_1 = g_0 \Big/ \left( 1 + \frac{W_3}{W_2} g_0 \ln \widetilde{\Delta} \right)$$

$$\approx g_0 \Big/ \left( 1 + \frac{W_3}{W_2} g_0 \ln|g_0| \right),$$

which differs from the parquet form[6] only in replacing $g_0$ by $g_1$.

Region *4* makes a nonperturbative contribution that has been discussed in detail in Refs. 5 and 6. It is obtained by substituting Eqs. (30) and (43) in Eq. (8), summing over $K$ by the saddle-point method, and summing over $N$ with the aid of the formula

$$\text{Im} \sum_{N=N_0}^{\infty} \Gamma(N+b) a^N (g_0 - i0)^N f(N) = \frac{\pi}{(ag_0)^b}$$

$$\times \exp\!\left(-\frac{1}{ag_0}\right) f\!\left(\frac{1}{ag_0}\right), \tag{46}$$

$$ag_0 > 0, \quad N_0 \gg 1,$$

which is valid for slowly varying functions $f(N)$. It is obtained by expanding $f(N)$ in a Fourier integral, using Eq. (90) of Ref. 6, and including only the long wavelength Fourier components. The arbitrariness in determining $f(N)$ for $N < N_0$ makes it possible to satisfy the condition of slow variation for any function $f(N)$ which does not vary more rapidly exponentially. The unusual phenomenon associated with the divergence of the series is that the sum in Eq. (46) is determined by arbitrarily large $N$ (so the result is independent of $N_0$), but the value of $f(N)$ for a finite $N = 1/ag_0$ appears in the result. Thus, the correction factor which distinguishes the exact $A_N^K$ from the Lipatov asymptote (23), is determined by Eqs. (30) and (43), and is negligible for $N \to \infty$, will lead to a substantial difference between the complete nonperturbative contribution and that calculated from the Lipatov asymptote (see Eq. (133) of Sec. 7.7 for $M = 1$ below):

$$[\Sigma(0,\kappa)]_{\text{nonpert}} \equiv i\Gamma_0(\kappa^2) = [\Sigma(0,\kappa$$

Approximating the series (17) by the sum of the contributions (45) and (47) and substituting in Eq. (14), we have

$$\kappa_0^2 - \kappa_c^2 = \kappa^2 \left[ 1 + W_2 g_1 \ln \frac{\Lambda}{\kappa} \right]^{1/4} + i\Gamma_0(\kappa^2),$$

$$\kappa^2 = -E - i\Gamma, \qquad (48)$$

which is solved similarly to Eq. (93) of Ref. 6 and determines the relationship between the damping $\Gamma$ and the renormalized energy $E$ with bare energy $E_B = -\kappa_0^2$ in the parametric form:

$$\Gamma = \Gamma_c e^x \sin \varphi, \quad E = -\Gamma_c e^x \cos \varphi,$$

$$-E_B + E_c = \Gamma_c e^x (4K_4 |g_1| x)^{1/4} \{\cos(\varphi + \varphi/4x) - \tan(\varphi/3) \sin(\varphi + \varphi/4x)\}, \qquad (49)$$

where $E_c$ is given by Eq. (108) of Ref. 6,

$$\Gamma_c = \Lambda^2 \exp\left\{ -\frac{1}{4K_4 |g_1|} \right\}, \qquad (50)$$

and $x(\varphi)$ is a unique function in the interval $0 < \varphi \pi$ analogous to that shown in Fig. 2 of Ref. 6 and given by (cf. Eq. (100) of Ref. 6)

$$\sin\left(\varphi + \frac{\varphi}{4x}\right) = B \cos\left(\frac{\varphi}{3}\right) \frac{e^{-4x/3}}{x^{1/4}} I\left(\frac{x}{x_0}\right), \qquad (51)$$

in which the constant $B$ is equal to

$$B = \sqrt{\pi} c_2 \left(\frac{3}{4}\right)^{1/4} \left(\frac{\widetilde{\Delta}}{3K_4 |g_0|}\right)^{7/4} \left(\frac{4}{3} x_0\right)^{7/6} \sim \left(\ln \frac{1}{|g_0|}\right)^{7/3}. \qquad (52)$$

The minimum values of $\Delta$ and $x$ are attained simultaneously and, to logarithmic accuracy, are

$$\Delta_{\min} \equiv \widetilde{\Delta} \approx \frac{21}{4} K_4 |g_0| \ln \frac{1}{|g_0|}, \quad x_{\min} \approx \frac{7}{4} \ln \ln \frac{1}{|g_0|}. \qquad (53)$$

The minimum distance to the false pole[3] is of order $|g_0| \ln \ln(1/|g_0|)$ and the "effective interaction" turns out to be logarithmically weak for small $\kappa$.

## 5. THE QUASIPARQUET APPROXIMATION

Calculating the density of states requires knowledge of the self energy $\Sigma(p,\kappa)$ for finite momenta.[6] As when $p=0$, this quantity consists of nonperturbative and quasiparquet contributions. The nonperturbative contribution will be important only at large negative $E$, where it is determined directly by the Lipatov asymptote and is given by Eq. (132) of Sec. 7.7 with $M=1$, $n=0$, and $p_1 = -p_2 = p$. For $p$ below some $p_0$, $[\Sigma(p,\kappa)]_{\text{nonpert}}$ is independent of $p$, while for $p \gtrsim p_0$ it falls off rapidly with increasing $p$. Given the logarithmic accuracy of the subsequent calculations (Sec. 8, Ref. 6), the following result is adequate:

$$[\Sigma(p,\kappa)]_{\text{nonpert}} \approx [\Sigma(0,\kappa)]_{\text{nonpert}} \Theta(p_0 - p),$$

$$p_0 \sim \kappa \left( \frac{1}{|g_0|} \ln \frac{1}{|g_0|} \right)^{1/2}. \qquad (54)$$

Major difficulties arise in calculating the quasiparquet contribution to $\Sigma(p,\kappa)$, which corresponds to a logarithmic approximation of arbitrary finite order including the principal asymptote in $N$ of the expansion coefficients. The principal logarithmic approximation for calculating $\Sigma(p,\kappa)$ requires knowledge of the four-tail vertex $\Gamma^{(0,4)}(p,k,q)$, which depends on three substantially different momenta, $p \gg k \gg q \gg \kappa$.[6] The method used above allows us to find the quasiparquet contribution to $\Gamma^{(0,4)}$ for $p \sim k \sim q \gg \kappa$. Writing an expansion of the type (8),

$$\Gamma^{(0,4)}(p,p,p) = \sum_{N=1}^{\infty} g_0^N \sum_{K=0}^{N-1} A_N^K \left( \ln \frac{\Lambda}{p} \right)^K \qquad (55)$$

(with different coefficients $A_N^K$) and noting that $\Gamma^{(0,4)}$ satisfies an equation like (15) with $V(g_0) \equiv -2\eta(g_0)$, in place of Eq. (20) we obtain

$$-KA_N^K = \sum_{M=1}^{N-K} [W_{M+1}(N-M) + V_M] A_{N-M}^{K-1},$$

$$K = 1, \ldots, N-1, \qquad (56)$$

which, given that $V_1 = 0$,[10] gives the following instead of Eqs. (27) and (28):

$$A_N^{N-K} = -2(-W_2)^N \frac{1}{(K-1)!}$$

$$\times \frac{(-W_3)^{K-1}}{(-W_2)^{2K-1}} (N \ln N)^{K-1}, \quad K \sim 1. \qquad (57)$$

The quasiparquet contribution to the sum (55), which is determined by the coefficients (57), has the form

$$[\Gamma^{(0,4)}(p,p,p)]_{\text{quasiparq}}$$

$$= -2 g_0 \Bigg/ \left( \Delta + \frac{W_3}{W_2} g_0 \ln \Delta \right) \Bigg|_{\Delta = 1 + W_2 g_0 \ln(\Lambda/p)}$$

$$\approx \frac{-2 g_1}{1 + W_2 g_1 \ln(\Lambda/p)}. \qquad (58)$$

In the principal logarithmic approximation, calculating $\Gamma^{(0,4)}(p,k,q)$ requires summation of the parquet graphs (Fig. 4a) obtained by successive splitting of simple vertices into two parts joined by two lines. When the order of the graph is increased by unity, the smallness $\sim g_0$ associated with the additional vertex is compensated by the large logarithm as-of two bricks and an irreducible four-tailed vertex (Fig. 4d). Setting up the equations of type $c$ and $d$ for the two other bricks, we obtain a system of 7 equations for 8 variables.[19] All the variables are uniquely determined by specifying the vertex $R^0$, so that

$$\Gamma^{(0,4)}(p,k,q) = F\{R^0(p,k,q)\}, \quad (59)$$

where $F\{\ldots\}$ is a functional. The parquet approximation corresponds to replacing the irreducible four-tail vertex with a simple vertex $R^0(p,k,q) = -2g_0$, so that the system $b$-$d$ then corresponds to the sum of the graphs $a$.

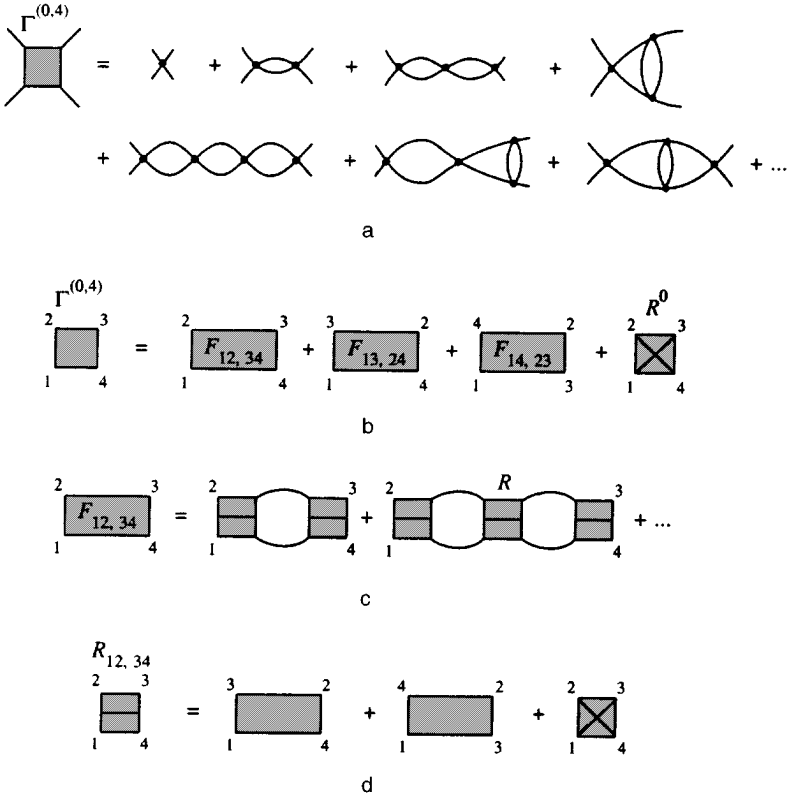

FIG. 4. (a) A parquet sequence of graphs for $\Gamma^{(0,4)}$ (obtained by successive splitting of simple vertices into two parts joined by a pair of lines); (b–d) the system of equations for the complete vertex $\Gamma^{(0,4)}$, three "bricks" $F_{ij,kl}$, three "crossed out" vertices $R_{ij,kl}$, and an irreducible vertex $R^0$. The approximation $R^0 = -2g_0$ corresponds to summing the sequence of diagrams $a$.

In Appendix 2 it is shown that the vertex $R^0$ depends only on the maximum momentum and Eq. (59) takes the form

$$\Gamma^{(0,4)}(p,k,q) = F\{R^0(p,p,p)\}. \quad (60)$$

Taking $k=q=p$ and reverting Eq. (60), we can obtain an approximation for $R^0$ that corresponds to the result (58), after which Eq. (60) solves the problem in principle.

The function $F\{\ldots\}$ is determined by the system of equations $b$-$d$ (Fig. 4), which is very complicated and has never been solved. Usually the approximation technique of Sudakov[20] as refined by Polyakov[21] is used. This method does not assume any specific approximation for $R^0$, but is based on logarithmic calculations of the type

$$\int_0^\Lambda \frac{k^3 dk}{(k^2+\kappa^2)^2} = \int_{\sim\kappa}^\Lambda \frac{k^3 dk}{k^4} = \ln\frac{\Lambda}{\kappa} + O(1). \quad (61)$$

The lower limit in the second integral is determined only to order of magnitude, but this uncertainty does not affect the larger logarithm and shows up only as a quantity of $O(1)$. At first glance, this method of calculation is justified only in the principal logarithmic approximation; in fact, the substitution $\kappa \to c\kappa$ in the principal logarithms (8) yields

$$A_N^N \left(\ln\frac{\Lambda}{\kappa}\right)^N \to A_N^N \left(\ln\frac{\Lambda}{\kappa} - \ln c\right)^N, \quad (62)$$

which leads to a change in the coefficients for the lower-order logarithms,

$$A_N^{N-K} \to A_N^{N-K} + \frac{(-\ln c)^K}{K!} N^K A_N^N, \quad K \sim 1. \quad (63)$$

According to Eq. (28), however, $A_N^{N-K} \sim A_N^N (N \ln N)^K$ and the principal asymptote in $N$ is "insensitive" to the substitution $\kappa \to c\kappa$. Thus, in the principal order in $N$, logarithmic calculations are permissible in an arbitrary finite logarithmic approximation. This makes it possible fully to employ the parquet scheme for calculating the quasiparquet contribution to $\Gamma^{(0,4)}$.

Substituting Eq. (58) in the parquet equation[19,21,11]

$$\Gamma^{(0,4)}(x,x,x) = R^0(x,x,x) + \frac{1}{2} K_4(n+8)$$

$$\times \int_0^x dt [\Gamma^{(0,4)}(t,t,t)]^2, \quad (64)$$

where $x = \ln(\Lambda/p)$, we obtain $p$

with increasing $R$, so that the derivative of $S(E,R)$ with respect to $R$ does not go to zero; thus, in the instanton method the variation of the Hamiltonian (1) with respect to $\varphi$, which leads to Eq. (71), does not vanish.

This difficulty can be overcome by minimizing $H\{\varphi\}$ for a fixed instanton radius $R$ followed by an essentially non-Gaussian integration with respect to $R$.

### 7.2. Expansion near an "anomalous" instanton

For the $(N-1)$th coefficient of the expansion of the $M$-point Green function (see Eqs. (29) and (58) of Ref. 6) we have

$$[G_M(x_1,\alpha_1,\ldots,x_M,\alpha_M)]_{N-1}$$

$$= Z_0(\kappa)^{-1} \int \frac{dg}{2\pi i} \int D\varphi \, \varphi_{\alpha_1}(x_1)\ldots\varphi_{\alpha_M}(x_M)$$

$$\times \exp\{-H\{\kappa,g,\varphi\} - N \ln g\}, \qquad (78)$$

where the Hamiltonian $H\{\kappa,g,\varphi\}$ is given by Eq. (1) and

$$Z_0(\kappa) = \int D\varphi \, \exp[-H\{\kappa,0,\varphi\}] = \left(\prod_s \frac{2\pi}{\lambda_s^0 + \kappa^2}\right)^{n/2} \qquad (79)$$

(the $\lambda_s^0$ are the eigenvalues of the operator $\hat{p}^2$). Let us introduce three expansions of unity under the integral of Eq. (78),

$$1 = \left(\int d^4x |\varphi(x)|^4\right)^4 \int d^4x_0$$

$$\times \prod_{\mu=1}^{4} \delta\left(-\int d^4x |\varphi(x)|^4 (x-x_0)_\mu\right),$$

$$1 = \int d^4x |\varphi(x)|^4 \int_0^\infty d \ln R^2$$

$$\times \delta\left(-\int d^4x |\varphi(x)|^4 \ln\left(\frac{x-x_0}{R}\right)^2\right). \qquad (80)$$

$$1 = \int d^n u \, \delta(\mathbf{u} - \mathbf{v}\{\varphi\}),$$

where the unit vector $\mathbf{v}$ is fixed by the condition

$$\mathbf{v}\{\varphi\} \| \int d^4x |\varphi(x)|^l \varphi(x), \quad l > 1. \qquad (81)$$

Making the change of variables

$$x - x_0 = R\tilde{x}, \quad \varphi_\alpha(x_0 + R\tilde{x}) = R^{-1} \tilde{\varphi}_\alpha(\tilde{x}) \qquad (82)$$

and changing the order of integration, we have, on dropping the tildes,

$$[G_M]_{N-1} = \int_0^\infty d \ln R$$

$$-\frac{1}{2}\sum_s (\lambda_s^L+\kappa_R^2)(C_s^L)^2 -\frac{1}{2}\sum_s\sum_\alpha (\lambda_s^T+\kappa_R^2)(C_s^{T,\alpha})^2\Bigg\},$$

where

$$\omega_s = g_c \int d^4x\, \varphi_c^3(x) e_s^L(x),$$

$$\sigma_s = \int d^4x\, e_s^L(x)[-\Delta\varphi_c(x) + \kappa_R^2\varphi_c(x) + g_c\varphi_c(x)^3],$$

$$\gamma_s^0 = 4\int d^4x\, \varphi_c^3(x) e_s^L(x) \ln x^2,$$

$$\gamma_s^\mu = 4\int d^4x\, \varphi_c^3(x) e_s^L(x) x_\mu, \quad \mu=1,2,3,4,$$

$$\beta_s = \int d^4x\, \varphi_c^l(x) e_s^T(x). \tag{89}$$

Transforming the $\delta$-functions in the exponent using the formula

$$\delta(x) = \frac{1}{2\pi}\int_{-\infty}^\infty d\tau\, e^{i\tau x} \tag{90}$$

and reducing the expression in the exponent to a sum of squares, for the integral with respect to $g$, $C^L$, and $C^T$ in Eq. (88) we obtain the result

$$\frac{g_c}{(2\pi)^{(n+5)/2}}\left(\prod_s \frac{2\pi}{\lambda_s^L+\kappa_R^2}\right)^{1/2}\left(\prod_s \frac{2\pi}{\lambda_s^T+\kappa_R^2}\right)^{(n-1)/2}$$

$$\times(\langle\beta^2\rangle_T)^{-(n-1)/2}\left(\prod_{\mu=1}^4 \langle\gamma^\mu\gamma^\mu\rangle\right)^{-1/2} [N\langle\gamma^0\gamma^0\rangle + \langle\omega^2\rangle$$

$$\times\langle\gamma^0\gamma^0\rangle - \langle\omega\gamma^0\rangle^2]^{-1/2} \exp\Bigg\{-N\ln g_c - H\{\kappa_R,g_c,\varphi_c\}$$

$$+\frac{1}{2}\langle\sigma^2\rangle - \frac{1}{2}\frac{\langle\sigma\omega\rangle^2}{N+\langle\omega^2\rangle}$$

$$-\frac{1}{2}\frac{[\langle\sigma\gamma^0\rangle(N+\langle\omega^2\rangle) - \langle\sigma\omega\rangle\langle\omega\gamma^0\rangle]^2}{(N+\langle\omega^2\rangle)[\langle\gamma^0\gamma^0\rangle(N+\langle\omega^2\rangle) - \langle\omega\gamma^0\rangle^2]}\Bigg\}, \tag{91}$$

where the following notation has been introduced:

$$\langle fg\rangle = \sum_s \frac{f_s g_s}{\lambda_s^L+\kappa_R^2}, \quad \langle fg\rangle_T = \sum_s \frac{f_s g_s}{\lambda_s^T+\kappa_R^2}. \tag{92}$$

The average $\langle fg\rangle$ is expressed in terms of the Green function of the operator $\hat{M}_L + \kappa_R^2$, which, given the obvious symmetry, can be used to prove that (as already used in Eq. (91)) the following quantities vanish:

$$\langle\sigma\gamma^\mu\rangle = \langle\omega\gamma^\mu\rangle = 0, \quad \mu\neq 0; \quad \langle\gamma^\mu\gamma^{\mu'}\rangle = 0, \quad \mu\neq\mu', \tag{93}$$

For a ''regular'' instanton that satisfies Eq. (71) with $\kappa\to\kappa_R$, the coefficients $\sigma_s\equiv 0$ and the exponent of Eq. (88) contains no terms linear in the deviations. The insolubility of Eq. (71) means that the coefficients $\sigma_s$ cannot be taken equal to zero. For a correct transition from (88) (83) to Eq. (91) they must be sufficiently small that the shifts in the variables $C_s^L$ during diagonalization of the quadratic form do not take the expansion (88) beyond its limits of applicability. The question arises as to how to choose $\varphi_c$ so as to ensure the required smallness of $\sigma_s$. When the instanton is chosen in accordance with Section 7.3, the sum $\Sigma_s\sigma_s C_s^L$ turns out to be proportional to the sum $\Sigma_s\gamma_s^0 C_s^L$, whose magnitude is fixed at zero owing to the presence in Eq. (88) of an appropriate $\delta$-function. In that way, terms linear in the deviations are eliminated exactly from the exponent of Eq. (88).

### 7.3. Choice of instanton

We choose the function $\varphi_c(x)$ by minimizing $H\{\kappa_R,g_c,\varphi\}$ with the additional condition (85) fixing the instanton radius. This yields

$$-\Delta\varphi_c(x) + \kappa_R^2\varphi_c(x) + g_c\varphi_c^3(x) + \mu\varphi_c^3(x) \ln x^2 = 0 \tag{94}$$

($\mu$ is a Lagrange multiplier[23]) with which it is easy to show that

$$\sigma_s = -\frac{\mu}{4}\gamma_s^0, \quad 2\omega_s - \frac{\mu}{4}\gamma_s^0 = (\lambda_s^L+\kappa_R^2)$$

$$\times\int d^4x\, \varphi_c(x) e_s^L(x), \quad H\{\kappa_R,g_c,\varphi_c\} = N, \tag{95}$$

from which we have

$$\langle\sigma^2\rangle = \left(\frac{\mu}{4}\right)^2\langle\gamma^0\gamma^0\rangle, \quad \langle\sigma\omega\rangle = -\frac{\mu}{4}\langle\omega\gamma^0\rangle,$$

$$\langle\sigma\gamma^0\rangle = -\frac{\mu}{4}\langle\gamma^0\gamma^0\rangle, \quad \langle\omega^2\rangle = -2N + \frac{\mu}{8}\langle\omega\gamma^0\rangle,$$

$$\langle\omega\gamma^0\rangle = \frac{\mu}{8}\langle\gamma^0\gamma^0\rangle, \tag{96}$$

with which Eq. (91) is greatly simplified and Eq. (88) takes the form

$$[G_M]_{N-1} = \left(\int d^4x\, \varphi_c(x)^4\right)^5 \int d^4x_0 \int_0^\infty d\ln R^2$$

$$\times \int d^n u\, \delta(|\mathbf{u}|-1) u_{\alpha_1}\ldots u_{\alpha_M} R^{-4-M}$$

$$\times\left(\int d^4x\, \varphi_c^{l+1}(x)\right)^{n-1} \varphi_c\left(\frac{x_1-x_0}{R}\right)$$

$$\ldots\varphi_c\left(\frac{x_M-x_0}{R}\right)\left(\frac{D_0}{D_T}\right)^{(n-1)/2}$$

$$\left(-\frac{D_0}{D_L}\right)^{1/2} g_c(2\pi)^{-(n+5)/2} N^{-1/2}$$

$$\times\prod_{\mu=0}^4 \langle\gamma^\mu\gamma^\mu\rangle^{-1/2}(\langle\beta^2\rangle_T)^{-(n-1)/2}$$

$$\times\exp(-N-N\ln g_c), \tag{97}$$

where the following notation for the determinants has been introduced:

$$D_L = \prod_s (\lambda_s^L + \kappa_R^2), \quad D_T = \prod_s (\lambda_s^T + \kappa_R^2),$$

$$D_0 = \prod_s (\lambda_s^0 + \kappa_R^2). \tag{98}$$

### 7.4. Explicit form of the instanton

The substitution $\mu = (-g_c)\mu_0$ and the transition to the function $\phi_c(x)$ in accordance with the first of Eqs. (72) eliminates $g_c$ from Eq. (94). For small $\kappa_R$, when $\mu_0 \sim \kappa_R^2 \ln \kappa_R$ (see below), Eq. (94) shares two scale lengths: a length $|x| \sim 1$ which determines the localization radius of the instanton ''core'' and a length $|x| \sim \kappa_R^{-1}$ to which the instanton ''tail'' extends. This makes it possible to carry out the analysis in two overlapping regions, $|x| \ll \kappa_R^{-1}$ and $|x| \gg 1$, and match the solutions.

*The region* $|x| \ll \kappa_R^{-1}$. For $\kappa_R = 0$ and $\mu = 0$, Eq. (94) has the exact solution (92) in which the solution (85) fixes the choice $R = 1$. Treating the terms containing $\kappa_R^2$ and $\mu$ as a perturbation, we have

$$\phi_c(x) = \frac{2\sqrt{2}}{z+1}\left[1 + \frac{1-z}{1+z} v(z)\right]_{z=x^2},$$

$$v(z) = \frac{\kappa_R^2}{4} \int_0^z dz \frac{(1+z)^4}{(1-z)^2 z^2}\left[-\ln(1+z) + \frac{z+2z^2}{(1+z)^2}\right]$$

$$+ \mu_0 \frac{z}{z-1}\left[-\ln z - \frac{z}{6} + \frac{3}{2}\right]. \tag{99}$$

Calculating the asymptote $v(z)$ for $z \gg 1$ and including only the terms that increase with $z$, for the region $1 \ll |x| \ll \kappa_R^{-1}$ we have

$$\phi_c(x) = \frac{2\sqrt{2}}{x^2}\left\{1 + \frac{1}{2} \kappa_R^2 x^2 \ln|x| + \left[\frac{1}{6}\mu_0 - \frac{3}{4}\kappa_R^2\right] x^2\right.$$

$$\left. + 3\kappa_R^2 \ln^2|x| + \left[2\mu_0 - \frac{11}{2}\kappa_R^2\right]\ln|x| - \frac{1}{x^2}\right\}.$$

$$\tag{100}$$

*The region* $|x| \gg 1$. When the terms that are nonlinear in $\phi_c(x)$ are neglected, Eq. (94) has the solution $B|x|$

the hydrogen atom. The substantially different $\mu_s$ cover the range of values $s(s+1)/6$ with a degree of degeneracy $s(s+1)(2s+1)/6$ ($s=1,2,\ldots$). For the determinants in Eq. (114) we obtain

$$-\bar{D}_R(1) = 2\exp\left\{\frac{15}{2} + \sum_{s=3}^{\infty} \frac{s(s+1)(2s+1)}{6}\right.$$
$$\times\left[\ln\left(1 - \frac{6}{s(s+1)}\right) + \frac{6}{s(s+1)}\right.$$
$$\left.\left.+ \frac{18}{s^2(s+1)^2}\right]\right\} \approx 578,$$

$$\bar{D}_R\left(\frac{1}{3}\right) = \exp\left\{\sum_{s=2}^{\infty} \frac{s(s+1)(2s+1)}{6}\left[\ln\left(1 - \frac{2}{s(s+1)}\right)\right.\right.$$
$$\left.\left.+ \frac{2}{s(s+1)} + \frac{2}{s^2(s+1)^2}\right]\right\} \approx 0.872. \quad (123)$$

### 7.7. Vertex part and nonperturbative contribution

For the integral over $d^n u$ in Eq. (113) we have[16]

$$\int d^n u\, \delta(|\mathbf{u}|-1) u_{\alpha_1}\ldots u_{\alpha_{2M}} = \frac{2\pi^{n/2}}{2^M \Gamma(M+n/2)} I_{\alpha_1\ldots\alpha_{2M}}, \quad (124)$$

where $I_{\alpha_1,\ldots,\alpha_{2M}}$ represents the sum of the terms $\delta_{\alpha_1\alpha_i}\ldots\alpha_{\alpha_j\alpha_k}$ obtained by all possible pairings (in the following we exclude $I_{\alpha_1,\ldots,\alpha_{2M}}$ from the definition of $G_{2M}$). Using the algebra of factorial series[6] it is easy to show that the transition to the vertex part of $\Gamma^{(0,M)}$ proceeds according to the formula

$$[\Gamma^{(0,M)}(p_1,\ldots,p_M)]_N = [G_M(p_1,\ldots,p_M)]_N$$
$$\times (p_1^2 + \kappa^2)\ldots(p_M^2 + \kappa^2). \quad (125)$$

Taking the Fourier transform of Eq. (113) and using the relation

$$\langle \phi_c \rangle_p = \frac{\langle \phi_c^3 \rangle_p - \mu_0 \langle \phi_c^3 \ln x^2 \rangle_p}{p^2 + \kappa_R^2} \approx \frac{\langle \phi_c^3 \rangle_p}{p^2 + \kappa_R^2} \quad (126)$$

for the Fourier components, which follows from the instanton equation (94), we have

$$[\Gamma^{(0,2M)}(p_1,\ldots,p_{2M})]_N$$
$$= c\,\frac{2\pi^{n/2}}{2^M\Gamma(M+n/2)}\left(\frac{4}{I_4}\right)^{M+5/2}\left(-\frac{4}{I_4}\right)^N$$
$$\times \Gamma\left(N + \frac{2M+n+4}{2}\right)\int_0^\infty d\ln R^2 R^{-4+2M}$$
$$\times \langle \phi_c^3 \rangle_{R_{P_1}}\ldots\langle \phi_c^3 \rangle_{R_{P_{2M}}}$$
$$\times \exp\left\{\frac{n+8}{3}\ln(\Lambda R) - Nf(\kappa R)\right\}. \quad (127)$$

Calculating the integral to logarithmic accuracy, for $p_i = 0$ we obtain

$$[\Gamma^{(0,2M)}(p_i=0,\kappa)]_N = c_{2M}(\ln N)^{-\gamma}\Gamma\left(N + \frac{n+8}{3}\right)$$
$$\times \left(-\frac{4}{I_4}\right)^N \kappa^{4-2M} \exp\left(\frac{n+8}{3}\ln\frac{\Lambda}{\kappa}\right),$$

$$c_{2M} = c\left(\frac{4}{I_4}\right)^{M+5/2}\left(\frac{2}{3}\right)^\gamma (I_3)^{2M}\,\frac{2\pi^{n/2}\Gamma(\gamma)}{2^M\Gamma(M+n/2)},$$

$$\gamma = M + \frac{n-4}{6}. \quad (128)$$

For $\kappa = 0$ and $p_i = p$, and noting that

$$\langle \phi_c^3 \rangle_p = 8\sqrt{2}\pi^2 p K_1(p), \quad (129)$$

we have

$$[\Gamma^{(0,2M)}(p)]_N = \tilde{c}_{2M}\Gamma\left(N + \frac{n+2M+4}{2}\right)$$
$$\times \left(-\frac{4}{I_4}\right)^N p^{4-2M} \exp\left(\frac{n+8}{3}\ln\frac{\Lambda}{p}\right),$$

$$\tilde{c}_{2M} = c\,\frac{4\pi^{n/2}(8\pi^2)^{2M}}{\Gamma(M+n/2)}\left(\frac{4}{I_4}\right)^{M+5/2}$$
$$\times \int_0^\infty dy\, y^{4M-5+(n+8)/3} K_1(y)^{2M}. \quad (130)$$

We define the nonperturbative contribution to the vertex $\Gamma^{(0,2M)}$ as

$$[\Gamma^{(0,2M)}]_{\text{nonpert}} = \sum_{N=N_0}^{\infty} [\Gamma^{(0,2M)}]_N (g_0 - i0)^N, \quad N_0 \gg 1. \quad (131)$$

The imaginary addition to $g_0$ originates in the need for analytic continuation from positive $g_0$ to negative,[2] which for $\text{Im}\,\kappa^2 < 0$ extends through the lower half plane. We obtain the nonperturbative contribution for the Lipatov asymptote by substituting Eq. (127) in Eq. (131), eliminating the imaginary part of $\kappa$ from the sum over $N$ by making the substitution $b \to be^{i\varphi}$, and summing in accordance with Eq. (90) of Ref. 6:

$$[\Gamma^{(0,2M)}(p_1,\ldots,p_{2M})]_{\text{nonpert}}^{\text{Lipatov}}$$
$$= i\pi c\,\frac{2\pi^{n/2}}{2^M\Gamma(M+n/2)}\left(\frac{4}{I_4}\right)^{M+5/2}\left(\frac{I_4}{4|g_0|}\right)^{M+(n+4)/2}$$
$$\times \exp\left(-\frac{I_4}{4|g_0|}\right)\int_0^\infty d\ln R^2 R^{-4+2M}\langle \phi_c^3 \rangle_{R_{P_1}}\cdots$$
$$\times \langle \phi_c^3 \rangle_{R_{P_{2M}}}\exp\left\{\frac{n+8}{3}\ln(\Lambda R) - \frac{I_4}{4|g_0|}f(\kappa R)\right\}.$$
$$(132)$$

For $p_i = 0$ this yields

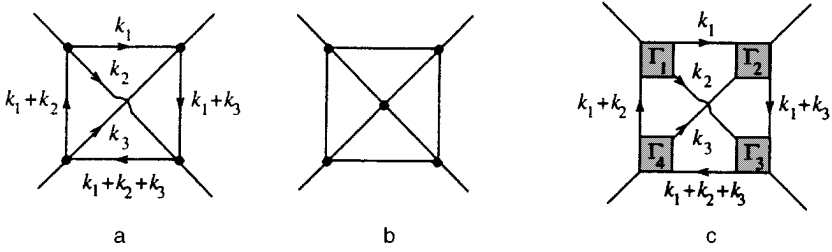

FIG. 5. Examples of skeleton (a,b) and nonskeleton (c) diagrams for a vertex $R^0$. The arrangement of the internal momenta in graphs $a$ and $c$ corresponds to zero internal momentum.

$$[\Gamma^{(0,2M)}(p_i=0,\kappa)]^{\text{Lipatov}}_{\text{nonpert}} = i\pi c_{2M}\left(\frac{I_4}{4|g_0|}\right)^{(n+8)/3}$$

$$\times \exp\left(-\frac{I_4}{4|g_0|}\right)\left(\ln\frac{1}{|g_0|}\right)^{-\gamma}\kappa^{4-2M}$$

$$\times \exp\left(\frac{n+8}{3}\ln\frac{\Lambda}{\kappa}\right). \quad (133)$$

The vertex $\Gamma^{(0,2)}$ coincides with the self energy and Eqs. (128) and (133) imply Eqs. (21) and (47). Equation (130) with $M=2$ can be used to recover the results of Refs. 14 and 16 for the expansion coefficients for the Gell-Mann–Low function (see Appendix 1).

The author thanks the participants in seminars at the Institute of Physical Problems and at the Physics Institute of the Academy of Sciences for their interest in this work.

This work has been supported financially by the International Science Foundation (Grants MOH 000 and MOH 300) and the Russian Fund for Fundamental Research (project No. 96-02-19527).

## APPENDIX 1

### Interrelation of the various normalizations

The definitions of the Gell-Mann–Low functions for the $\Lambda$- and $\mu$-renormalizations are[10]

$$W^{(\Lambda)}(g_0) = \left.\frac{\partial g_0}{\partial \ln \Lambda}\right|_{g_\mu,\mu=\text{const}},$$

$$W^{(\mu)}(g_\mu) = \left.\frac{\partial g_\mu}{\partial \ln \mu}\right|_{g_0,\Lambda=\text{const}}, \quad (A1)$$

and the relation between the renormalized and bare charges is determined by logarithmic expansions like (54),

$$g_\mu = \sum_{N=1}^{\infty} B_N g_0^N = \sum_{N=1}^{\infty} g_0^N \sum_{K=0}^{N-1} A_N^K \left(\ln\frac{\Lambda}{\mu}\right)^K, \quad (A2)$$

$$g_0 = \sum_{N=1}^{\infty} \widetilde{B}_N g_\mu^N = \sum_{N=1}^{\infty} g_\mu^N \sum_{K=0}^{N-1} \widetilde{A}_N^K \left(\ln\frac{\mu}{\Lambda}\right)^K, \quad (A3)$$

where $B_1 = \widetilde{B}_1 = 1$. The function $g_\mu(g_0,\Lambda/\mu)$ satisfies an equation like (15) with $V(g_0)\equiv 0$, and the coefficients $A_N^K$ in Eq. (A2) satisfy Eqs. (55) with $W_M \equiv W_M^{(\Lambda)}$ and $V_M \equiv 0$. $W_N^{(\mu)}$ and $\widetilde{A}_N^K$ are related in the same way. By finding the relationship between $\widetilde{A}_N^K$ and $A_N^K$ for small $N$, it is easy to confirm that the coefficients $W_N^{(\Lambda)}$ and $W_N^{(\mu)}$ coincide for $N=2,3$ but differ in the higher orders. For large $N$ the coefficients $B_N$ are defined by Eq. (130) with $p=\mu$ and $M=2$, since by definition $-2g_\mu = \Gamma_R^{(0,4)}(\mu) = Z^2\Gamma^{(0,4)}(\mu)$ (the coefficient $-2$ is introduced because of the condition $B_1=1$) and a $Z$ factor does not have to be included for large $N$ because of the slower rise in its coefficients. Reversion of the factorial series (A2) yields

$$\widetilde{B}_N \approx -\sum_{M=0,1,2,\ldots} B_{N-M}\frac{(N\widetilde{B}_2)^M}{M!} \approx -B_N \exp\left(-\frac{B_2}{a}\right), \quad (A4)$$

for large $N$ and this makes it possible to determine the $\widetilde{A}_N^K$. Given that $A_2^1 = -K_4(n+8)$, from Eq. (55) we have

$$W_N^{(\mu)} \approx -\widetilde{A}_N^1 - W_2^{(\mu)} N\widetilde{A}_{N-1}^0$$

$$\simeq \frac{n+8}{6} \widetilde{c}_4 \Gamma\left(N+\frac{n+8}{2}\right) a^N \exp\left(-\frac{A_2^0}{a}\right), \quad (A5)$$

and the substitution $A_2^0 = -(n+8)K_4 K_4(1-\ln(4/3))/2$ reproduces the results of Ref. 14 and 16, where the definitions of $g_\mu$ and $g_0$ differ by a factor of 3!

## APPENDIX 2

### Properties of the irreducible vertex $R^0$

Let us introduce the three-momentum notation $R^0(p,k,q)$ for the irreducible vertex $R^0(p_1,p_2,p_3,p_4)$ with $\kappa=0$, where

$$2p = p_4 - p_3, \quad 2k = p_1 - p_2, \quad q = p_1 + p_2 = -p_3 - p_4, \quad (A6)$$

and the numbers of the ends are chosen so that $p \geq k \geq q \geq 0$. In general, when $p \geq k \geq q$ the following logarithmic expansion is valid for $R^0$,

$$R^0(p,k,q) = \sum_{N=1}^{\infty} g_0^N \sum_{\substack{K,L,M,\\K+L+M\leq N-1}} A_N^{K,L,M}\left(\ln\frac{\Lambda}{p}\right)^K$$

$$\times \left(\ln\frac{\Lambda}{k}\right)^L \left(\ln\frac{\Lambda}{q}\right)^M \quad (A7)$$

and in order to prove that the vertex $R^0$ depends on only the maximum momentum, it is sufficient to establish its finiteness for $k=q=0$ and $p \neq 0$. Then $p_4 = p$, $p_3 = -p$, and $p_1 = p_2 = 0$ and it can be said that the momentum $p$ "passes" from vertex 4 to vertex 3.

The proof is based on the fact that it is necessary to distinguish skeleton graphs of type $a$ and $b$ (Fig. 5) from nonskeleton graphs $c$ obtained from the skeleton graphs by replacing the simple vertices by four-tail blocks. For the case of the contribution of graph $a$,

$$R^0_{(a)}(p_i=0) \sim g_0^4 \int d^4k_1 \int d^4k_2 \int d^4k_3$$
$$\times G_{k_1}G_{k_2}G_{k_3}G_{k_1+k_2}G_{k_1+k_3}G_{k_1+k_2+k_3}, \quad (A8)$$

it is easy to see that the skeleton graphs for the zero external momenta contain only singly logarithmic divergences originating from the region of integration where all the internal momenta are of the same order of magnitude and are small. The absence of other divergence is related to the impossibility of finding a single momentum $k_i$ that determines the arguments of the two $G$-functions, two momenta $k_i$ and $k_j$ that determine the arguments of the four $G$-functions, etc. When momentum $p$ passes through any pair of vertices, the singly logarithmic divergences are cut off at momentum $p$, since it enters the argument of at least one $G$-function. In this way, the graph is finite for $k=q=0$. The divergences of nonskeleton graphs are cut off just as are the divergences of the skeleton graphs from which they are obtained.

We proceed to formalize the remarks. A graph with $E$ external lines in the $N$th order of perturbation theory has $I=2N-E/2$ internal lines and $L=N-E/2+1$ independent integrations (loops), and diverges for $d-4$ at large momenta as $k^{4-E}$.[10] In the following we are concerned with four-tail graphs ($E=4$) with $L$ loops, $2L$ internal lines, and $L+1$ logarithmically divergent vertices.

*Definition.* A graph is a skeleton graph if it is impossible to choose $L'$ independent integrations in it such that $L'<L$ and the integration momenta $k_1,k_2,\ldots,k_{L'}$ fully determine the arguments of the $2L'$ or more Green functions.

*Lemma 1.* A skeleton graph contains only singly logarithmic divergences, which are removed when the momentum $p$ passes through any pair of vertices.

The proof is evident from the definition and explanations given after Eq. (A8).

*Lemma 2.* In a skeleton graph it is impossible to isolate a subdiagram with the form of a four-tail block.

Assume, on the contrary, that this is possible. In the isolated block there will be $M<L$ independent integrations with respect to the momenta $k_1,k_2,\ldots,k_M$ which will enter only in the $2M$ Green functions corresponding to the internal lines of the block. The remaining $L'=L-M$ integration momenta will fully determine the arguments of the $2L-2M=2L'$ Green functions, which contradicts the definition of a skeleton diagram.

*Lemma 3.* In a graph which is not a skeleton diagram it is always possible to isolate a subgraph in the form of a four-tail block.

In a nonskeleton graph there are $L'$ integration momenta which completely determine the arguments of the $2L'$ Green functions. The remaining $M=L-L'$ integration momenta enter only in $2L-2L'=2M$ Green functions. The corresponding $2M$ internal lines will, in general, lie in separate clusters. Let us isolate one such cluster (the $i$-th) containing $N_i$ vertices, $L_i$ loops, and $I_i$ lines and remove it from the diagram, breaking $E_i$ lines. Obviously, $I_i=2N_i-E_i/2$ and $L_i=N_i-E_i/2_1$, so that

$$I_i-2L_i=E_i/2-2. \quad (A9)$$

Obviously, all the $E_i$ are even. The cases $E_i=0$ and $E_i=2$ correspond to disconnected diagrams and self-energy inserts and are assumed to be excluded. Then $E_i \geq 4$ and, in view of Eq. (A9), we have

$$I_i-2L_i \geq 0. \quad (A10)$$

Summing over all $i$ and noting that $\Sigma_i I_i=2M$ and $\Sigma_i L_i=M$, we see that the equality holds in Eq. (A10), so that (see Eq. (A9)) $E_i=4$ and all the removed clusters are four-tailed. Since at least one cluster exists, the assertion is proven.

*Lemma 4.* A nonskeleton graph can be obtained from some skeleton graph by replacing all or part of the simple vertices with four-tail blocks.

By successively isolating the four-tail blocks in a nonskeleton graph and replacing them with simple vertices, we ultimately arrive at a skeleton graph. Following this procedure in opposite order, we obtain a proof of the assertion.

*Lemma 5.* A nonskeleton graph of order $N$ obtained from a skeleton graph of order $M$ contains a logarithmic divergence of order no greater than $N-M+1$, which is cut off by passage of momentum $P$ through any pair of vertices.

Let us write down as an example the contribution of the graph $c$ (Fig. 5) obtained from graph $a$:

$$R^0_{(B)}(p_i=0) \sim \int d^4k_1 \int d^4k_2 \int d^4k_3 G_{k_1}G_{k_2}G_{k_3}G_{k_1+k_2}$$
$$\times G_{k_1+k_3}G_{k_1+k_2+k_3}\Gamma_1(0,-k_1,-k_2,k_1+k_2)$$
$$\times \Gamma_2(0,k_1,k_3,-k_1-k_3)\Gamma_3(0,k_1+k_3,k_2,-k_1-k_2$$
$$-k_3)\Gamma_4(0,-k_1-k_2,-k_3,k_1+k_2+k_3).$$

The divergence originates from the region $k_i \sim k$, for which arguments of all the vertices are $\sim k$:

$$R^0_{(B)}(p_i=0) \sim \int \frac{dk}{k} \Gamma_1(k)\Gamma_2(k)\Gamma_3(k)\Gamma_4(k).$$

The block $\Gamma_i$ containing $N_i$ simple vertices has a maximum divergence $\sim (\ln \Lambda/k)^{N_i-1}$ if it is included in the parquet sequence of Fig. 4a. Since $\Sigma_i N_i=N$, a graph with $M$ blocks diverges no more rapidly than $(\ln \Lambda/k)^{N-M+1}$, so the divergence is eliminated when momentum $p$ passes through any pair of vertices. There are no other divergences. If not all the momenta are small ($\sim k$), then a graph diverges exponentially as $k \to 0$ and suppresses the logarithmic divergences of the blocks.

Thus, all the graphs are finite for $k=q=0$ and $p \neq 0$, and $R^0(p,k,q)$ depends only on the maximum momentum $p$.

---

[1] The renormalized energy $E$ enters into expressions of the type (5).[13]
[2] The functions $W$ and $\eta_2$ are calculated in Ref. 10 for $\mu$-renormalization and differ from the $\Lambda$-renormalization used here, but have the same leading coefficients (see Appendix 1); the definitions of charge and the Gell-Mann–Low function differ in Ref. 10 by a factor of $6K_4$.
[3] In Refs. 14 and 16 the coefficients $W_N$ have been calculated in a

$\mu$-renormalization, for which this assumption is not correct (see Appendix 1).

4) This can be done following the arguments of Section 4.6 of Ref. 6, using the most general form of the counterterms (see Eqs. (3.26)–(3.28) of Ref. 10).

Translated by D. H. McNeill